\documentclass[11pt]{article}

\usepackage{times}
\usepackage{tikz}
\usepackage{cite}
\usepackage{hyperref}
\usepackage{setspace}

\usepackage{siunitx}  
\DeclareSIUnit{\kilocalorie}{kcal}
\DeclareSIUnit{\bit}{bit}

\usepackage{graphicx}

\newenvironment{sciabstract}{%
\begin{quote} \bf}
{\end{quote}}

\title{Control across scales: signals, information, and adaptive biological mechanical function}

\author
{James Clarke$^{1 \ast}$, Jake McGrath$^{1 \ast}$, Colin Johnson$^{1}$, Jos\'{e} Alvarado$^{1 \ast \ast}$\\
\\
\normalsize{$^{1}$Center for Nonlinear Dynamics, Department of Physics,} \\[0.5ex] 
\normalsize{The University of Texas at Austin. Austin, TX, USA}\\\\
\normalsize{$^\ast$: Co-first authors; these authors contributed equally to this work.}\\
\normalsize{$^{\ast \ast}$ : To whom correspondence should be addressed; E-mail: alv@chaos.utexas.edu}
}

\date{}

\begin{document} 

\maketitle

\begin{sciabstract}
\singlespacing
Biological systems perform an astonishing array of dynamical processes---including development and repair, regulation, behavior and motor control, sensing and signaling, and adaptation, among others. Powered by the transduction of stored energy resources, these behaviors enable biological systems to regulate functions, achieve specific outcomes, and maintain stability far from thermodynamic equilibrium. These behaviors span orders of magnitude in length and time: from nanometer-scale molecular motors driving morphogenesis to kilometer-scale seasonal migrations, and from millisecond reflexes to millennia of evolutionary adaptations. While physical laws govern the dynamics of biological systems, they alone are insufficient to fully explain how living systems sense, decide, adapt, and, ultimately, control their dynamics. In this article, we argue that control theory provides a powerful, unifying framework for understanding how biological systems regulate dynamics to maintain stability across length and time scales far from equilibrium.
\end{sciabstract}

\section*{Introduction}

Objects undergoing motion have inspired and awed humans. Examples include the unpredictable trajectory of a curveball, the reliable take-off of a commercial airplane, or the remarkable stability of a rolling bicycle. These examples are well described by the laws of physics. Additional examples of motion come from life itself. Individual cells can change their external shape, rearrange their internal structure, divide, crawl, and pull on the environment. Collections of cells can migrate together, heal wounds, curve planar tissues, lengthen tissue regions, and ultimately sculpt themselves to form well-structured embryos. Entire organisms, too, can walk, run, jump, stand, fly, and balance. These kinds of mechanical tasks in biology are not just awe-inspiring: they are necessary for survival. To which extent do the laws of physics help in understanding functional mechanical behavior in biology?

Biological mechanics is ultimately and inherently a non-equilibrium process. At the molecular scale, active, chemomechanical processes convert free energy reserves to mechanical work. One major example of such processes is given by molecular motors (e.g. myosin, kinesin, and dynein), which hydrolyze adenosine triphosphate (ATP) to generate sliding motion within the cytoskeleton, a network of biological polymers (e.g. actin and microtubules). How do motors on $\sim\qty{100}{\nano\meter}$ length scales drive mechanics on cellular and organismal scales $(\sim\qty{10}{\micro\meter}-\qty{1}{\meter})$? Condensed matter physics has seen remarkable progress in reducing macroscopic phenomenological observables—such as the speed of sound in air, the elastic modulus of a solid, or the conductivity of a thin-film heterostructure—to models of interactions between microscopic constituents \cite{chaikin_principles_1995}.

\begin{figure}[htbp]
    \centering
    \includegraphics[width=0.95\textwidth]{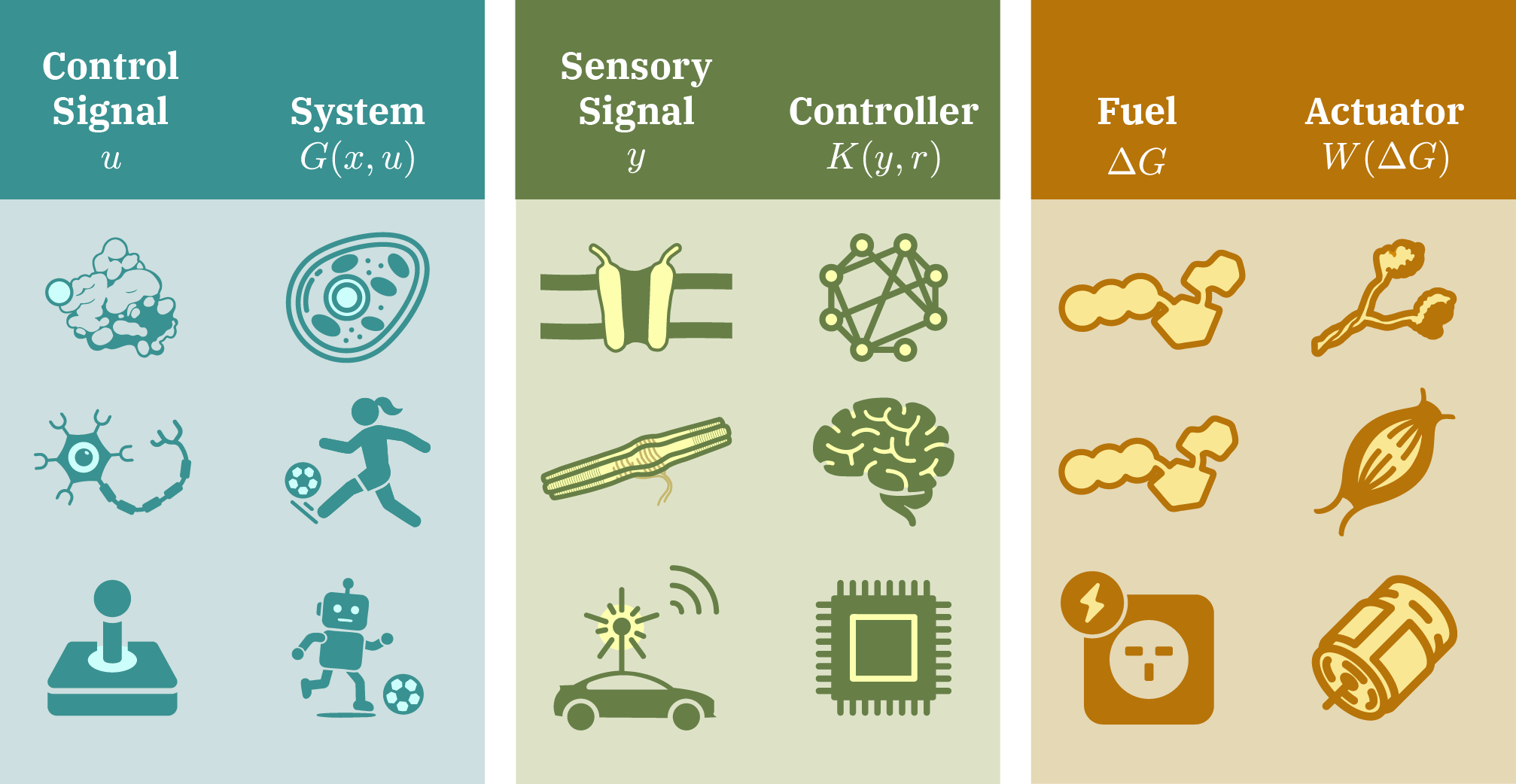}
    \caption{Essential constituents of functional mechanical behavior. \textbf{Top row}: Animal cells. \textbf{Center row}: Organisms. \textbf{Bottom row}: Robots. \textbf{From top left to bottom right}: examples of constituents: protein phosphorylation, cells, stretch-activated ion channels, network of signaling proteins, adenosine triphosphate (ATP), myosin; neurons, humans, spindle fibers, the central nervous system, ATP, muscle; potentiometers, robots, LiDAR, microprocessors, electric power outlets, direct-current (DC) motors.}
    \label{fig:figure1}
\end{figure}

Despite these successes, one look at a soccer match demonstrates there is still a long way to go in connecting microscopic-scale activity to macroscopic-scale mechanics in biology. Can the laws of physics predict how a striker decides to move their body to slip past an opponent and score a goal? There are at least two major hurdles that hinder such an approach. The first challenge lies in the complex hierarchical structure found in living systems. Constituents are heterogeneous, protein-protein interactions can change their conformational and functional state (e.g. via post-translational modifications like phosphorylation), and several levels of spatial organization interact across scales. Many existing studies have focused on this challenge \cite{schaffer_mapping_2021}. A second challenge, perhaps less studied by physicists, lies in decision-making processes.

Cells and organisms continuously make decisions \cite{perkins_strategies_2009, schall_decision_2005}. These decisions include a combination of a response to present conditions, obtained through observation; and an execution of learned patterns, obtained through past experience and stored in memory. Furthermore, these decisions adapt. Individual organisms adapt in their own lifetime; populations adapt across generations; entire species adapt over evolutionary time scales. Therefore, understanding biological behavior requires understanding not just the physics of motion, but also the decision-making processes that effect motion. In organisms, decisions are executed via a variety of \emph{control signals}, which activate actuators, such as molecular motors. Which formalism describes these decisions? Control theory and information theory are quantitative frameworks that have attracted increasing interest among physicists. Control theory determines the control signal needed to produce a desired motion or behavior in engineered systems like robots and prosthetics. Information theory quantifies the processing and transmission of information-bearing signals.

Researchers have interestingly found success in describing complex mechanical function without control or information theory. Early studies on active matter coincided with interest in self-organization. Several studies demonstrated patterns and behaviors that resulted from a collection of individual active agents. Examples include molecular motors \cite{surrey_physical_2001}, bird flocks \cite{bialekStatisticalMechanicsNatural2012, viraghFlockingAlgorithmAutonomous2014}, bacterial swarms \cite{dombrowski_self-concentration_2004}, cell tissues \cite{friedlCollectiveCellMigration2009, alertPhysicalModelsCollective2020}, and human crowds \cite{dyer_consensus_2008}. Meanwhile, concepts from self-organization also helped in simplifying human and robotic locomotion. For example, compliance in human limbs provides dynamical stability to perturbation in the absence of control signals, i.e. “mechanical intelligence” \cite{blickhan_intelligence_2006}. Further, mechanical interactions and self-organization are key ingredients in “embodied intelligence” in biologically inspired and soft robotics \cite{pfeifer_self-organization_2007, laschi_soft_2025}. Sometimes, a simple mechanical model lacking complex control suffices. Human locomotion can be well described by minimal models containing a few effective parameters \cite{full_templates_1999}. Meanwhile for cellular locomotion, negative durotaxis can be explained by simple clutch models lacking specific stiffness sensing \cite{isomursu_directed_2022}.

Although purely mechanical descriptions of biological function can be sufficient in certain cases, information-bearing control signals ubiquitously prevail in cells and organisms and often cannot be neglected. This prevalence is reflected in several applications of control and information theory to biological systems, such as animal locomotion \cite{cowan_feedback_2014}, neuronal systems \cite{quian_quiroga_extracting_2009}, and bacterial chemotaxis \cite{bourret_molecular_2002}. These two theoretical frameworks not only apply to phenomenological-scale function in biology. Information theory has also become a core component of stochastic thermodynamics, which describes molecular-scale systems undergoing non-equilibrium fluctuations. Stochastic-thermodynamic studies have uncovered relationships between mechanical work, entropy, free energy, mutual information, and memory \cite{parrondo_thermodynamics_2015}. Furthermore, fundamental thermodynamic laws can limit the performance of feedback control systems \cite{bechhoeferControlTheoryPhysicists2021b}. Control-theoretical approaches have been applied to stochastic, active, and soft systems, as reviewed in several recent articles \cite{blaber_optimal_2023, takatori_feedback_2025, levine_physics_2023, alvarado_optimal_2025}. However, the fact that control and information theory have benefited descriptions of physical and biological systems at molecular and organismal scales is fascinating, and leads us to ask whether these widely disparate scales can be connected.

In this article, we argue that control and information theory can quantitatively connect microscopic-scale, non-equilibrium molecular activity and macroscopic-scale, phenomenological biological mechanical function. Furthermore, disparate timescales could be quantitatively connected as well, from task-specific to evolutionary. To this end, we first introduce two important concepts from control theory: feedforward and feedback control, especially in the context of molecular-scale activity and biological mechanical function. Next, we review molecular-scale stochastic thermodynamics and information theory, as well as macroscopic-scale experiments probing animal behavior using control theory and robophysics. We then extend these concepts to increasingly longer timescales, from population dynamics to adaptation and evolution. Finally, we review applications to synthetic cells, robotics, and prosthetics. We anticipate that further research in these areas will help elucidate how molecular activity helps drive adaptive mechanical behavior in biology.

\section*{Control signals influence system dynamics}

Living systems continuously consume chemical free energy to remain out of equilibrium. However, flows of energy are never constant in cells and organisms. Energetic flows are regulated by upstream control signals that regularly activate and deactivate energy consumption and activation. Control signals prevail in biology at multiple scales. In intracellular contexts, biochemical regulation is carried out by post-translational modification of proteins and with small signaling molecules like calcium and cAMP \cite{berridgeVersatilityUniversalityCalcium2000,schwartzManyDimensionsCAMP2001}. In extracellular contexts, secreted molecules allow a cell to sample the collective state of its neighbors and bias its mechanics accordingly \cite{albertsMolecularBiologyCell2008}. Cytokines, chemokines, and morphogens diffuse or are actively transported through the extracellular matrix to form concentration gradients \cite{wolpertPositionalInformationPatterning2011,mullerExtracellularMovementSignaling2011,inomataScalingPatternFormations2017}. Binding of these proteins to surface receptors triggers different types of intracellular signaling cascades \cite{kholodenkoCellsignallingDynamicsTime2006}. At organismal scales, endocrine bursts serve as whole-body set-points\cite{lightmanCrucialRolePulsatile2010}, while motor neurons dispatch millisecond commands to muscle groups to actuate limbs \cite{delvecchioHumanCentralNervous2019}.

In non-muscle cells, pulsatile RhoA/Rac signalling and $\mathrm{Ca}^{2+}$ flashes activate ROCK and myosin-light-chain kinase, phosphorylating myosin-II and locally stiffening the actomyosin cortex before any load is felt \cite{bementActivatorInhibitorCoupling2015,graesslExcitableRhoGTPase2017,michauxExcitableRhoADynamics2018}. In skeletal muscle the process is analogous but faster and more ordered: a motor-neuron action potential depolarises the sarcolemma, the sarcoplasmic reticulum releases $\mathrm{Ca}^{2+}$, and $\mathrm{Ca}^{2+}$ binding to troponin slides tropomyosin aside to expose myosin-binding sites on actin, triggering cross-bridge cycling in microseconds \cite{berchtoldCalciumIonSkeletal2000}. Both contractile mechanisms are a direct consequence of activation by control signals.

What causes control signals to activate or deactivate actuation? Control theory offers a framework to answer this question (Fig. 1a). An essential component is the \textit{controller}, denoted $K$, which issues \textit{control signals} $u(t)$ to the physical system $G$ (Fig.~\ref{fig:figure1}). The system undergoing actuation is governed by a dynamical rule that depends not only on the state $x(t)$ of the system (as is often the case in dynamical systems), but also the control signal $u(t)$. Among the simplest examples of controllers is the \textit{feedforward controller}, which produces a predetermined control signal $u(t)$ (Fig.~\ref{fig:figure2}, left). Examples of control that rely heavily on feedforward mechanisms include traffic lights, lawn sprinklers, circadian rhythms, and intracellular signaling cascades.

Feedforward control (a.k.a. “open loop control”) has certain advantages. Since $u(t)$ is predetermined, one can more easily predict how a system will behave in the future. When coupled to positive feedback, feedforward can store memory  \cite{bechhoeferControlTheoryPhysicists2021b}. Feedforward also provides the foundation for more sophisticated forms of control, including optimal, adaptive, and acausal control
\cite{sutton_reinforcement_2018, alvarado_optimal_2025, takatori_feedback_2025}. In particular, optimal control methods combine knowledge of a system, usually through a predictive model, in determining a sequence of control signals to yield a desired functionality. A recent theoretical advance by \textbf{Shankar et al.} \cite{shankar_design_2024} further builds on optimal control, developing symmetry-based activity patterns to steer topological defects in active liquid crystals.

Although control signals steer dynamics, biological systems rarely rely entirely on feedforward alone. Its major drawback is that a predetermined control signal $u(t)$ does not adapt to changing circumstances. To overcome this limitation, control systems, including living systems, often incorporate sensors.

\section*{Sensors convey information to decision-making processes}

Life is equipped with a wide array of sensors that sample mechanical, chemical, and inertial cues across scales. At the molecular level, mechanosensitive proteins such as talin and $\alpha$-catenin unfold under load, exposing cryptic binding sites that trigger Rho-GTPase and Src–FAK signaling \cite{yaoMechanicalResponseTalin2016,delrioStretchingSingleTalin2009,renCrossregulationsTwoConnected2024,yonemuraACateninTensionTransducer2010}. Stretch-activated ion channels, such as Piezo, TREK, and TRP, convert membrane tension into $\mathrm{Ca}^{2+}$ or $\mathrm{Na}^{+}$ influx within microseconds \cite{costePiezo1Piezo2Are2010,ranadeMechanicallyActivatedIon2015,wangSingleMoleculeFRET2014}. Cells aggregate these point detectors into organ-level arrays: otolith hair-cell bundles report linear acceleration \cite{hudspethHowEarsWorks1989}, cochlear hair cells transduce acoustic vibrations \cite{coreyKineticsReceptorCurrent1983}, spindle fibers in muscle register properties like length and velocity \cite{kandel_principles_2000}, cutaneous Ruffini and Pacinian endings encode shear and vibration \cite{johnsonRolesFunctionsCutaneous2001}, and mechanosensitive primary cilia bend to sense fluid flow in the kidney \cite{nauliPolycystins122003}.

Mechanosensing is hard-wired into the cytoskeleton itself: force-dependent catch bonds in actin–myosin networks \cite{guoMechanicsActomyosinBonds2006,veigelLoaddependentKineticsForce2003}, strain-stiffening in intermediate filament lattices \cite{janmeyViscoelasticPropertiesVimentin1991, kreplakBiomechanicalPropertiesIntermediate2007}, and load-modulated microtubule dynamics \cite{dogteromForceGenerationDynamic2005,dogteromActinMicrotubuleCrosstalk2019} couple strains from applied stresses back to RhoA activation and downstream transcription factors \cite{dupontRoleYAPTAZ2011}. At the tissue level, muscle spindles sense when a muscle is stretched and quickly send signals through nerves to the spinal cord, which adjusts muscle activity in real time—even within a single step \cite{proskeProprioceptiveSensesTheir2012}.

Sensors “close the loop” left open by feedforward control, sending \textit{sensory signals} $y(t)$ back into biochemical and neural controllers (Fig. 1b). A second fundamental control scheme is \textit{feedback control}, where the controller $K$ takes in the sensory signal $y(t)$ and produces a new control signal $u(t)$ (Fig.~\ref{fig:figure2}, center-left). Feedback controllers adjust the control signal in order to minimize the difference between the measured signal and a desired reference value. Examples include a car with cruise-control engaged, maintaining a reference speed; cells adjusting myosin activity to hold tension constant across a tissue \cite{stamenovic_tensional_2020}; the knee-jerk reflex that restores muscle length after a tendon tap \cite{kandel_principles_2000}; and impedance-controlled robots that vary limb actuation to maintain a virtual mechanical impedance \cite{hogan_adaptive_1984} (Fig.~\ref{fig:figure1}).

Feedback has certain advantages over feedforward: it corrects for noise and disturbances. Further, feedback control does not require detailed knowledge of the system’s dynamics, relying instead on sensory signals. However, experimentalists working with feedback control need to tune certain parameters through trial and error or heuristic algorithms. Furthermore, feedback can introduce instabilities under certain conditions. Because of these advantages and disadvantages, biological decision-making processes incorporate signaling cascades that incorporate both feedback and feedforward mechanisms. Inside many cells, mutually activating and inhibiting modules — Rho/ROCK, RAC/PAK, and YAP/TAZ circuits — integrate calcium pulses, tension‐dependent talin unfolding, and membrane curvature to modulate actomyosin contractility in real time \cite{michauxExcitableRhoADynamics2018,humphreyMechanotransductionExtracellularMatrix2014a,dupontRoleYAPTAZ2011,debellyInterplayMechanicsSignalling2022,somlyoCa2SensitivitySmooth2003,mitchellVisceralOrganMorphogenesis2022b}. In tissue and whole-organ contexts, interneuron ensembles perform the same logic in electrical form: peripheral reflex arcs couple stretch-receptor input to $\alpha$ motor neuron output within 20–40 ms \cite{soteropoulosLonglatencyResponsesMechanical2020}.

\section*{Control over microscopic activity: lessons from stochastic thermodynamics}

Chemical free energy powers many instances of functional mechanical behavior across spatial scales in living systems. As we have seen, the non-equilibrium activity of molecular motors is one common mechanism that converts free energy to mechanical work (Fig.~\ref{fig:figure1}). In order to understand energy transduction, we first consider biological energetic flows. We will see that energy transduction does not provide a complete description: entropy production is perhaps as important, in light of control signals and information processing.

Consider the energetic flows of human metabolism. An average adult human consumes $\sim\qty{2000}{\kilocalorie/\day}$, which corresponds to a Gibbs free energy consumption rate $\dot G = \qty{100}{\watt}$ per human, or $\qty{3}{\pico\watt}$ per cell, on average \cite{milo_cell_2015}. This chemical free energy is the fuel for the power stroke of molecular motors. One could therefore seek to quantify the thermodynamic efficiency $\eta=\Delta W/\Delta G$ of a human being. To answer this question, consider where most people begin and end their day: their homes. Net work against gravity is zero; efficiency is zero; and all energy is ultimately dissipated. Meanwhile, the fact that efficiency is zero over long enough timescales has been known in human locomotion studies, which instead consider the cost of transport $\mathrm{COT} = \frac{\Delta G}{\Delta x}$, a physical quantity that measures the free energy dissipated per unit distance $\Delta x$ traveled \cite{alexander_models_2005}.

Despite the above arguments, energetic demands are still important considerations in biological mechanics. On short, task-specific timescales, instantaneous efficiency can be nonzero. For example, when extending one’s arms during weightlifting or one’s legs during running, human muscle’s efficiency is approx. 25\% \cite{barclay_chapter_2019}. And although virtually all energy is ultimately dissipated, energetic demands still contribute to selective pressures over evolutionary timescales. A recent theoretical study by \textbf{Davis et al.} \cite{davis_active_2024} of active systems further demonstrates the importance of timescales. Slow, quasistatic driving minimizes dissipation due to state transitions, but increases the total time (and hence total energy) over which active processes dissipate. An optimal driving timescale minimizes the sum of these two contributions, in contrast to passive systems where slow, quasistatic driving minimizes dissipation.

So far energy dissipation appears to be an undesirable byproduct. Could there be benefits to dissipation in mechanical tasks? Given that an average human outputs $\qty{100}{\watt}$ of heat, the rate of entropy generation $\dot S = \dot Q/T$ is approximately $\qty{0.3}{\watt/\kelvin}$ (with temperature $T = \qty{300}{\kelvin}$), or $\qty{10}{\femto\watt/\kelvin}$ per cell. Entropy generation is synonymous with an increase in phase-space volume \cite{sethnaStatisticalMechanicsEntropy2021}. In the case of heat output, the increase is largely incurred in the environment, resulting in irreversibility \cite{tanScaledependentIrreversibilityLiving2021}. Viewing life as an entropy-maximizing process \cite{kleidon_life_2010} provides a non-equilibrium analog to the second law of thermodynamics \cite{martyushev_maximum_2006}. Notice that maximizing entropy appears to stand at odds with minimizing energy dissipation. This apparent contradiction demonstrates that optimization considerations are helpful tools in formulating and solving problems, but care must be taken when extending interpretations to living systems.

Comparing the space of physical configurations to the space of control signals provides a connection between thermodynamics and information theory. An early study by Landauer demonstrated that logically irreversible processes, such as destroying one bit of memory, produces heat, with corresponding minimum entropy generation of $S = k \log 2$ \cite{landauer_irreversibility_1961}. Here, one bit refers to the conventional units of the Shannon entropy, which is defined in analogy to thermodynamic entropy \cite{shannon_mathematical_1948}. One interpretation of Shannon entropy is the number of possible signals one could transmit. Although Landauer’s argumentation was motivated by computer memory, studies of biological information processing have used Landauer’s bound as a starting point \cite{ouldridge_thermodynamics_2017}.

Motivated by these studies, we now approximate the maximum mutual information extraction rate of an average human cell. Given the $\qty{10}{\femto\watt/\kelvin}$ from above, assuming an upper bound $\dot S/k = \dot I/k_s$ (with $k$ Boltzmann’s constant and $k_s=\log 2~\mathrm{bit}$), we find approx. $\dot I = \qty{e9}{\bit/\s}$. Do cells really process a gigabit of information per second? Although the answer is likely “no”, it remains a frontier of biophysics research to quantify information flows and answer this question definitively. Since Landauer’s study, information has gained traction as a physical quantity \cite{georgescu_60_2021}. More recently, attention has turned toward descriptions of \textit{bipartite systems}, which formalize exchanges of energy, work, and information between two coupled, stochastic systems \cite{barato_information-theoretic_2013, horowitz_thermodynamics_2014}. These formalisms help in describing simple systems such as optically trapped beads \cite{jun_high-precision_2014} and ATP synthase motors \cite{grelier_unlocking_2024}. It would be interesting to continue to extend these fundamental descriptions of molecular-scale, non-equilibrium active systems to macroscopic mechanical function.

\section*{Control over macroscopic mechanical behavior: lessons from robophysics}

Although molecular scale activity ultimately powers many instances of biological mechanical function on cellular and organismal scales, challenges remain in connecting such widely separated scales. Despite these challenges, control theory continues to successfully offer quantitative descriptions of mechanical function in cells and organisms. Concepts from information and control theory suffice to describe how a system can regulate its components to perform coordinated tasks. Control theory has been applied to describe cellular memory \cite{jiang_memorizing_2021}, adaptive immune response \cite{rahman_importance_2018}, insect flight \cite{chatterjee_integration_2022,cowan_feedback_2014}, and gait patterns \cite{lam_contribution_2006,pearson_generating_2004}, among others—all of which rely on integrated control strategies to maintain robust mechanical function. In addition to developing control-theoretical models, the recent field of robophysics has offered an additional experimental platform to study the intersection of biology and control theory experimentally.

In robophysics experiments, researchers construct simplified, robotic, physical model systems that preserve key features of biological behavior. Despite their strengths, robophysical model systems have notable limitations. They are not biological in nature: actuators and controllers operate differently from muscles and cellular machinery. Consequently, robophysical models only approximate biological function. While useful for uncovering general principles, these models cannot fully replicate the molecular-scale processes that occur in cells and tissues.

Despite these limitations, there are several advantages of robophysical model systems compared to experiments with living organisms. First, robophysics systems abstract away the multi-scale complexity and molecular detail of a living organism and focus on its function. Feedback methods do not require knowing a system’s dynamics, relying instead on measurement and calibrated gains. Feedforward methods may rely on a model of the system’s dynamics, but simple, coarse-grained models with lumped parameters are often sufficient. Second, compared to live animals, robophysical model systems can be easily measured and controlled. For example, a recent robophysical study systematically investigated schools of robotic fish interacting with turbulent flows \cite{zhang_collective_2024}. Here, researchers were able to precisely control the spatial arrangement of fish within the school to determine the effects of different arrangements. Similarly, robophysical models of insect flight allow for direct manipulation of internal and external parameters—such as wing-beat cadence, joint configuration, and actuation power—enabling controlled exploration of mechanisms that are otherwise difficult to isolate in live organisms. This level of control has revealed key physical principles underlying biological function, such as supra-resonant wing-beat frequencies \cite{wold_supra-resonant_2025}, passive vibrational stabilization \cite{taha_vibrational_2020}, and transitions between flight modes \cite{gau_bridging_2023}. Robophysics experiments therefore overcome some of the difficulties in achieving repeatable and controlled experiments when working with living systems. Finally, robophysical models can access internal configurations that are often challenging to measure directly in living organisms. For instance, tracking force propagation and body configurations in robotic models has been key in studying gait patterns across different media \cite{aguilar_review_2016}. Specifically, manipulating tail and flipper configurations in robophysical models has clarified how aquatic animals use these appendages to locomote over soft, deformable terrain \cite{mcinroe_tail_2016,mazouchova_flipper-driven_2013}.

These advantages of robophysical systems have significantly deepened our understanding of locomotion across a wide range of organisms. Notably, complementary work by \textbf{Chong et al.} \cite{chong_multilegged_2023,chong_general_2022} has leveraged robophysics to uncover principles of biomechanical coordination and develop a general control framework for multi-legged locomotion in complex, confined environments. In these studies, the authors first propose a generalized control scheme for coordinating high-dimensional, serially connected legged robots by combining biologically inspired gait design with geometric mechanics. Second, they introduce a novel theoretical framework that treats locomotion over rugged terrain as a problem of information transmission, demonstrating that reliable transport can be achieved through spatial redundancy. Together, these studies provide both a theoretical basis for locomotion through complex terrain and a practical control strategy across diverse robot morphologies—a key step toward creating reliable, adaptable, and deployable robotic systems for real-world environments. Additional advances by the Goldman and Choset groups have explored how geometric principles and mechanical intelligence underpin effective locomotion in both limbless and legged organisms across scales and substrates—offering simplified control strategies and improved performance in complex environments \cite{rieser_geometric_2024,wang_mechanical_2023,chong_coordination_2021}. Moreover, advances in snake robotics have shed light on the role of frictional anisotropy in undulatory locomotion \cite{schiebel_mitigating_2020} and have helped elucidate how snakes navigate slippery or granular terrains \cite{marvi_sidewinding_2014}. These insights have also driven the development of all-terrain robotic systems \cite{liu_review_2021} and informed improvements in limbless robot performance through the modulation of wave patterns during actuation \cite{chong_frequency_2021}.

Perhaps the most profound benefit of robophysical model systems lies not just in the ability to measure internal configurations: one can also measure energy consumption. This allows robophysical model systems to quantitatively characterize energetic flows during mechanical tasks, allowing for a deeper connection between thermodynamics and function. Recent studies of robotic fish have investigated this connection. Body shape and stroke patterns affect thrust and efficiency \cite{xie_experimental_2020}, and the configuration of fish within a school can disrupt vortices in turbulent flows and thus increase efficiency \cite{zhang_collective_2024}. Furthermore, a robophysical system that mimicked the nonlinear contraction of actin-myosin systems revealed how nonlinearity governs a tradeoff between energetic efficiency and power output in muscle actuation \cite{mcgrathHilltypeBioinspiredActuation2022,mcgrath_microscale_2025}. While robophysical models are not biological in nature---their actuators and controllers are often fueled by electric as opposed to chemical free energy---back-drivable, direct-current, electromagnetic motors exhibit some similarities to muscle undergoing concentric contraction, and their energetics exhibit similar trends under certain circumstances \cite{mcgrath_microscale_2025}. We anticipate that robophysical approaches will provide a powerful and tractable experimental platform that will test relationships between thermodynamics, control, and information, and how these relationships support mechanical function.

While control theory and robophysics offer powerful tools for probing the physical underpinnings of biological behavior at the organismal scale, the principles of control are equally valuable in understanding how biological systems operate across population and evolutionary timescales. In this broader context, control theory provides a framework for analyzing how feedback, regulation, and adaptation shape the dynamics of evolving populations and ecological interactions.

\section*{Adaptation and evolution}

Over time, living systems adapt their internal organization and behavior in response to changing environmental conditions. A broad body of interdisciplinary research on adaptive systems \cite{lansing_complex_2003} has modeled social-ecological systems \cite{preiser_social-ecological_2018}, the biosphere \cite{levin_ecosystems_1998}, cancer \cite{schwab_cancer_1996}, and immune networks \cite{iwasaki_control_2015}. Because of their capacity for self-reorganization, adaptive systems naturally lend themselves to analysis through machine learning techniques \cite{gheibi_applying_2021}. When adaptation is absent, a previously robust system could become unable to maintain stability once the environment changes. When adaptation is too aggressive, systems can become fragile, overcompensating for change. This tension raises a central question: how do systems adapt effectively to slow environmental shifts while remaining robust to temporary disturbances?

\begin{figure}[htbp]
    \centering
    \includegraphics[width=0.95\textwidth]{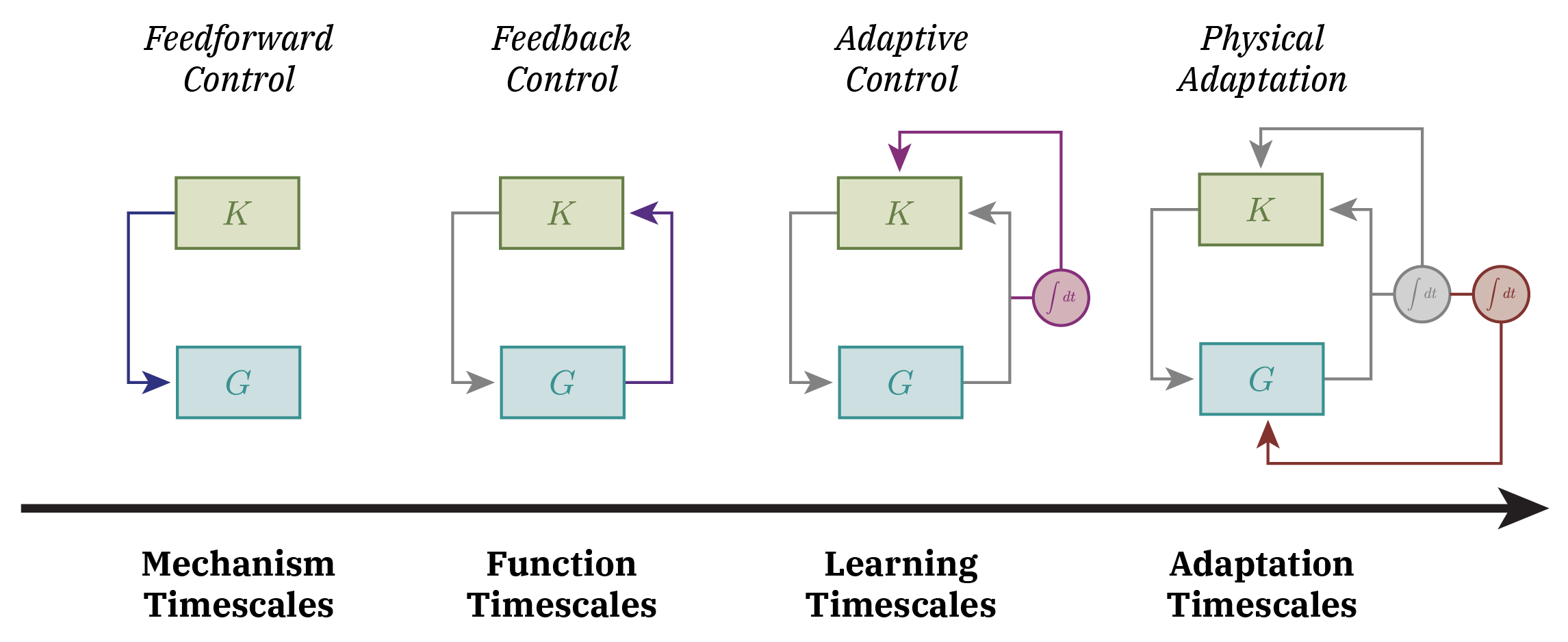}
    \caption{Control strategies establish mechanical function over distinct timescale ranges. \textbf{Left}: Feedforward control occurs over the timescales inherent to the physical mechanism through which control signals from the controller $K$ influence the system $G$. \textbf{Center-left}: Feedback control includes transmission of sensory signals from the system back to the controller, incurring additional information-processing time. \textbf{Center-right}: Adaptive control methods modify the parameters of the controller over long periods to accommodate gradual changes in the system. \textbf{Right}: Biological systems ultimately adapt over individual, population, and evolutionary timescales.}
    \label{fig:figure2}
\end{figure}

One quantitative approach to this problem is found in adaptive control methods \cite{bechhoeferControlTheoryPhysicists2021b} (Fig.~\ref{fig:figure2}, center-right). While internal parameters (such as gains in feedback control) remain constant in standard controllers, in adaptive control, these parameters are dynamically adjusted in response to environmental or system changes. For example, adaptive control algorithms can account for system wear, damage, or degradation while maintaining output or efficiency. In these adaptive feedback architectures, robustness emerges from feedback mechanisms that allow a system to buffer perturbations while adjusting to long-term environmental trends; negative feedback, in particular, plays a crucial role in stabilizing behavior by damping fluctuations and resisting noise. Adaptive control methods allow study of adaptation in complex, high-dimensional nonlinear systems because it can drive a system toward stable behaviors—such as steady states or limit cycles—even in the presence of uncertainty or changing conditions \cite{sinha_adaptive_1990}. To prevent overreaction to transient disturbances, robust adaptive control ensures that the system responds primarily to slow, sustained changes while ignoring fast perturbations through signal filtering, bounded parameter updates, or modification rules that suppress adaptation to noise or short-term fluctuations \cite{lavretsky_robust_2024}. Adaptive control methods can therefore achieve both adaptability and robustness. While adaptive control has been widely studied in engineered systems, it remains underexplored in the context of adaptive materials and living systems.

One example where adaptive control could be helpful is in population dynamics. Population numbers among interacting species are subject to relatively short, intragenerational fluctuations due to effects like reproduction, predation, or resource availability. On longer, intergenerational timescales,  slow drifts can occur due to changing environments. Classical models of population dynamics—such as the Lotka-Volterra equations—effectively capture coarse-grained population behaviors through lumped parameters. This approach successfully recapitulates observed behaviors, including oscillations \cite{smith_stability_1973,mougi_evolution_2010,dornelas_landscape-induced_2021}, equilibrium points \cite{slobodkin_conditions_1955}, extinction events \cite{escudero_extinction_2004,gilpin_ecocultural_2016}, and even chaos \cite{hanski_population_1993}. However, these models often struggle to accurately predict real-world population dynamics \cite{cerini_predictive_2023,a_lee-yaw_species_2022,levine_why_2025}, potentially because they do not incorporate feedback or learning on model parameters explicitly. Recent efforts in conservation and ecological management implicitly use feedback-control-like principles; for example, adjusting harvesting rates \cite{xu_harvesting_2005,beddington_harvesting_1977} or introducing predators to curb overpopulation or invasive species \cite{hoddle_restoring_2004,solomon_natural_1949}. Given living systems’ ability to observe, learn and adapt to changing circumstances, population dynamics problems are natural candidates for analysis through the lens of adaptive control. We anticipate adaptive control could provide a natural framework to more accurately model and manage animal populations threatened by climate change, habitat loss, urbanization, and species introduction.

On even longer timescales, the Darwinian evolution of species faces a similar challenge: evolving systems must respond to selective pressures while remaining robust to short-term fluctuations. Crucially, organisms do not just respond to their environments—they also modify them, creating feedback loops that shape their own evolutionary trajectories \cite{pausas_feedbacks_2022}. These feedback loops, whether positive or negative, can generate stability, instability, or even chaos \cite{robertson_feedback_1991} producing complex evolutionary dynamics. This interplay of robustness and adaptability closely parallels the central concern of adaptive control. However, one major difference between biological adaptation and adaptive control is that living systems and species adapt to accommodate changing circumstances. Meanwhile, adaptive control can only learn to cope with system degradation, and an external mechanics (e.g. a repair technician) is needed to have the physical system regain its desired mechanical function. In this sense, it is remarkable how life incorporates adaption in both control strategies and its physical components.

\section*{Applications and outlook}
\subsection*{Medicine, synthetic cells, and medical devices}
Control theory provides a mathematical and conceptual framework for understanding how systems respond to control signals, maintain stability, and achieve desired outcomes despite internal or external disturbances. This is especially important in medicine, where functional regulation spans multiple scales---from molecular interactions to organismal behavior---all working to restore balance in response to disturbances. Biological function is remarkable in its ability to support survival across these widely separated levels, and control theory provides a framework for formalizing this multiscale robustness.

In medicine, this perspective enables us to model complex physiological processes, design interventions that restore or modulate function, and build systems---both natural and synthetic---that adaptively regulate themselves. On small scales, recent progress highlights synthetic cells as programmable platforms capable of sensing, processing, and responding to physiological signals \cite{blain_progress_2014}. Many essential life processes have been reconstituted within these fully engineer-able synthetic chassis \cite{gaut_reconstituting_2021} enabling the design of feedback-regulated systems for maintaining homeostasis, delivering targeted therapeutics, and modulating immune responses \cite{sato_synthetic_2022}. Advances in protein design offer the ability to generate proteins with desired functionality, which can be integrated into synthetic cells to expand their biochemical and regulatory capabilities \cite{frohn_protein_2025}. At larger scales, recent advances in robotic prosthetics and exoskeletons have significantly improved mechanical mobility and support for patients \cite{bogue_exoskeletons_2009,mendez_current_2021}. More recent developments have produced adaptive, sensor-integrated devices that interface with the nervous system enabling more natural motor control and restoring limb function with increasing precision \cite{song_continuous_2024}. Moreover, energy landscape models and attractor dynamics offer powerful tools for modeling neurological and psychiatric disorders \cite{regonia_modeling_2021,rolls_attractor_2021} by capturing how alterations in brain network stability and transition dynamics underlie conditions such as schizophrenia and depression. In this framework, medical interventions can be viewed as controllers that reshape these energy landscapes and attractor dynamics—disrupting maladaptive configurations and guiding the system back toward a stable, healthy state. By framing medical interventions as control problems, we gain predictive power, design flexibility, and robustness—positioning control theory a natural language for future perspectives in medicine.

\subsection*{Human engineering, space omics, and closed-loop control}
Everything we have discussed — actuators, sensors, feedforward and feedback loops — converges into an emerging discipline of "space omics" \cite{rutterAstronautOmicsImpact2024}. It is no longer science fiction to posit that humans are well-poised to become a designed species.  Dr. Chris Mason’s \emph{The Next 500~Years} argues that we have both the moral duty and the technical roadmap to engineer our genome, microbiome, and built environments so that we are able to thrive on planets bathed in $\ge 400\;\mathrm{mSv\,yr^{-1}}$ of radiation and $<0.4\,g$ gravity \cite{Mason2021Next500Years}. The process of iterative feedback is clear: measure the molecular consequences of a stressor, enact a repair, and repeat exactly the feedback architecture championed throughout this article.

Spaceflight is already the most extreme \textit{in vivo} perturbation lab ever built. The NASA Twins Study showed that $>10^{5}$ molecular features (genomic, epigenomic, transcriptomic, proteomic, metabolomic, immunologic) shift in concert during a year in orbit and relax on return \cite{garrett-bakelmanNASATwinsStudy2019}. New consortia have scaled those measurements to dozens of astronauts, providing the data backbone for model-based control of human physiology \cite{overbeyCollectionBiospecimensInspiration42024}. In the language of control theory, a multi-omics profile is the full state vector $x(t)$; single-cell atlases define the system output $y(t)$, and programmable editors (CRISPR, epigenetic writers, engineered probiotics) constitute control signals $u(t)$.

Recent advances in synthetic biology and systems medicine have demonstrated that intracellular feedback circuits—such as LOCKR protein cages implementing PID-like logic—can precisely regulate molecular set-points at the nanomolar scale \cite{langanNovoDesignBioactive2019}. Concurrently, hybrid computational pipelines are beginning to integrate transcriptomic, proteomic, and metabolomic data into genome-scale models, offering the potential to translate omics snapshots into actionable control variables for cells and tissues \cite{maniGenomicsMultiomicsAge2025}. These developments lay the groundwork for more sophisticated control schemes, such as model-predictive control (MPC) \cite{eslamiControlStrategyBiopharmaceutical2024}, which could in theory anticipate biological trajectories, optimize intervention strategies, and adaptively update recommendations as new data become available.

However, significant challenges remain before such closed-loop, adaptive frameworks can be robustly implemented in complex organisms or clinical settings. Biological systems are highly nonlinear and context-dependent, and current models often lack the fidelity needed for reliable prediction, especially at the whole-organism level. Real-time, high-resolution omics sensing \textit{in vivo} is still limited. The safe, rapid actuation of interventions — such as multiplexed gene editing or metabolic rewiring — remains largely experimental and faces substantial technical and ethical hurdles. Furthermore, defining cost functions that meaningfully capture health status, risk, and resource trade-offs is a complex and unresolved problem.

Despite these limitations, the convergence of synthetic biology, omics technologies, and control theory represents a promising direction for precision medicine, especially in extreme or rapidly changing environments.  Control theory is attractive framework due to the fact that we do not need knowledge of the system $G$ as long as we (i) observe enough of the state (multi-omics readouts) and (ii) apply corrective control signals frequently relative to the disturbance timescale (radiation hits, micro-g de-conditioning, circadian desynchronization). Human engineering thus becomes a regulation problem, not a one-shot redesign: long-duration crews would fly with an onboard omics-to-MPC pipeline that continuously steers towards a range of molecular states compatible with health and performance in space. Continued progress in biosensing, modeling, and safe actuation will be essential to realize the vision of adaptive, model-driven biological control. For now, these approaches should be viewed as a research frontier—one that is rapidly advancing, but not yet ready for routine application in humans.

\section*{Conclusions}
Biological systems perform complex behaviors across an extraordinary range of spatial and temporal scales. From nanometer-scale molecular motors to organism-scale locomotion and population-level adaptation, living systems dynamically regulate their internal states to maintain stability in response to perturbations in ways that defy simple physical descriptions. To understand how such robustness emerges, here we argue that control and information theory offer powerful frameworks for understanding the dynamical processes of living systems.

Control architectures---including feedforward, feedback, and adaptive loops---shape the flow of energy and information across length and time scales. These flows allow biological systems to operate far from thermodynamic equilibrium and support mechanical function. Sensory networks transmit diverse physical and chemical signals to decision-making modules which determine control signals that modulate biochemical and mechanical states. Then, sensory networks measure again and continue the process once more. This control loop stabilizes internal states in the presence of noise and uncertainty from the surrounding environment.

Looking forward, control theory provides a lens to study population dynamics, evolutionary feedback, and the engineering of synthetic biological systems. As real-time, multi-omics sensing and actuation technologies mature, the vision of closed-loop, programmable, biological control may soon be realized. Integrating control, information, and thermodynamics offers not only a new language for understanding living systems---but a design framework for building the next generation of adaptive, resilient, and intelligent systems.

\section*{Acknowledgments}
This research was primarily supported by the National Science Foundation CAREER Grant No. DMR-2144380. This research was supported in part by the National Science Foundation through the Center for Dynamics and Control of Materials: an NSF MRSEC under Cooperative Agreement No. DMR-2308817. This research was supported in part by grant NSF PHY-2309135 and the Gordon and Betty Moore Foundation Grant No. 2919.02 to the Kavli Institute for Theoretical Physics (KITP).

\bibliographystyle{unsrt}
\bibliography{biblio}

\clearpage

\end{document}